# Mean Diurnal Variations of Cosmic Ray Intensity as Measured by Andyrchi Air Shower Array (E ≥ 100 TeV) and Baksan Underground Scintillation Telescope (E ≥ 2.5 TeV)

V. A. Kozyarivsky, A. S. Lidvansky, V. B. Petkov, and T. I. Tulupova

*Institute for Nuclear Research, Russian Academy of Sciences, Moscow, 119312*
*Russia*

## Abstract

The results of measuring diurnal variations of extensive air showers (EAS) with E ≥ 100 TeV detected by the Andyrchi air shower array in 1995—2002 are presented. It is shown that solar diurnal wave of intensity is formed by pressure variations and Compton—Getting effect (due to the orbital motion of the Earth). The data on cosmic ray anisotropy of the Baksan Underground Scintillation Telescope (BUST) are also presented for E > 2.5 TeV covering the period 1982—1998. The diurnal sidereal wave measured by both Andyrchi and BUST is due to anisotropy of the cosmic rays.

We have analyzed the data on the intensity of extensive air showers with energy E ≥ 100 TeV detected by the Andyrchi air shower array in the period 1995 – 2002. The Andyrchi array [1] is located at an altitude of ~ 2000 m above sea level and is a part of experimental facilities of the Baksan Neutrino Observatory, Institute for Nuclear Research, Moscow.

The counting rate (15-min values) of four-fold coincidences of any out of 37 detectors of the array was analyzed, as well as atmospheric pressure and temperature of the near-earth air at the instant of data output.

Andyrchi is an EAS array and was not specially designed for continuous operation. In addition, during the period under consideration there were serious damages of the array due to thunderstorms with subsequent terminations for repairing. As a result, the total data set included only 780 days. For this period we have detected $6\cdot10^{8}$ events, which resulted in an overall statistical accuracy at a level of $4.1\cdot10^{-5}$. The averaged daily waves of intensity, atmospheric pressure, and temperature were formed from these data in three times: solar, sidereal, and anti-sidereal.

The data selection was made in the following way: we excluded from processing those 15-min intervals for which the deviations from the daily mean value exceeded ± 3.5 σ. The daily information was eliminated in the case if the intensity dispersion for a given day exceeded three errors of dispersion.

The diurnal wave of pressure turned out to be nonzero in both sidereal and anti-sidereal times, which made an additional contribution of the barometric effect to the diurnal waves of shower intensity in these times. A correlation analysis of the dependence of the counting rate of four-fold coincidences on the atmospheric pressure variation gave the following values of the correlation coefficients (K) and barometric coefficient (R):

*Table 1 .*

| Time | K | R  %/mb |
|------|---|---------|
| Solar | 0.81 $\pm$ 0.14 | $-$ 1.106 |
| Sidereal | 0.71 $\pm$ 0.18 | $-$ 1.201 |
| Anti-sidereal | 0.91 $\pm$ 0.09 | $-$ 1.061 |

In order to correct the diurnal waves for pressure we have taken R for anti-sidereal time, since we consider that the anti-sidereal wave originates solely due to pressure variation so that its R value represents the barometric dependence in the best way. The correlation analysis of pressure-corrected temperature waves with intensity waves has shown no correlations in all chosen times.

After the correction of daily waves of intensity for pressure we performed the Fourier analysis in order to derive the first harmonics of these waves. The following values are obtained for the first harmonics:

*Table 2.*

| Time | Amplitude (x $10^4$) | Phase (h) |
|------|---------------------|-----------|
| Solar | 2.58 $\pm$ 0.58 | 6.53 $\pm$ 0.86 |
| Sidereal | 3.66 $\pm$ 0.58 | 0.20 $\pm$ 0.61 |
| Anti-sidereal | 0.82 $\pm$ 0.58 | 9.59 $\pm$ 3.00 |

In the first approximation the diurnal wave $I(t)$ of intensity of cosmic rays measured by a wide-angle telescope can be represented in the following form:

$$I(t) = I_0 + i_0 Cos\,\delta_0\,Cos\,\delta_{eff}\,Cos(t - t_0)\,,$$

where $I_0$ is the mean value of intensity, $\delta_0$ is the declination of the source of anisotropy, $\delta_{eff}$ is the effective declination of the wide-angle telescope. The relative amplitude of such a wave is equal to

$$A = \xi Cos\delta_0 Cos\delta_{eff},$$

which in practice is almost always less than the true degree of anisotropy $\xi$. The cosine of the effective declination of an array ($\delta_{eff} \cong 54°$ for Andyrchi) indicates how much the amplitude of the measured wave is reduced relative to the anisotropy vector projection $\xi Cos\delta_0$. If we make a correction for this reduction assuming $\delta_{eff} = 0°$, then the parameters of the intensity wave in the solar time are as follows (Table 3).

*Table 3.*

| Time | Amplitude (x $10^4$) | Phase (h) |
|------|------|------|
| Solar | 4.39 $\pm$ 0.98 | 6.53 $\pm$ 0.86 |

These values coincided (both in amplitude and phase) with those expected for the Compton—Getting effect (~ $4.7 \cdot 10^{-4}$ and 6 h) due to orbital motion of the Earth with the velocity $v \cong 30$ km/s. The first harmonic of the sidereal wave corrected for $Cos\delta_{eff}$ corresponds to the projection of the sidereal anisotropy vector whose parameters are given in Table 4.

*Таблица 4.*

| Time | $\xi Cos\delta_0$ ( x $10^4$) | Фаза (R.A.) |
|------|------|------|
| Sidereal | 6.22 $\pm$ 0.98 | 0.53 $\pm$ 0.61 |

This result is somewhat different from the results obtained by us using the data of the Baksan Underground Scintillation Telescope (BUST) [2] at E $\geq$ 2.5 TeV. In this paper we included the BUST data for 1982—1998. The underground telescope detected the flux of muons with the energy E $\geq$ 220 GeV, arriving at the telescope from 8 selected directions [3]. The selection criteria were the same as described above for the case of Andyrchi data. The only difference in method is due to the fact that there is no barometric effect at large depths. Only the temperature variation in the upper layers of the troposphere has a considerable effect. After a correction for the atmospheric temperature effect [4] we obtained the values of anisotropy vector projection for all selected directions. These values are tabulated in Table 5. The data for all selected directions coincide within the errors, which is indicative of reliability of the measurements made with the BUST.

*Table 5.*

| Direction | $\xi Cos\delta_0$ ( x $10^4$) | Phase R.A. (h) |
|-----------|-------------------------------|----------------|
| 2 from 6 | 10.11 ± 0.72 | 1.84 ± 0.28 |
| 6 - 7 | 10.56 ± 1.34 | 1.97 ± 0.48 |
| 7 - 8 | 10.51 ± 1.28 | 2.04 ± 0.46 |
| 6-7-8 | 9.91 ± 1.50 | 2.04 ± 0.58 |
| 5-6-7-8 | 9.38 ± 1.52 | 1.89 ± 0.62 |
| "West" | 11.16 ± 1.74 | 1.50 ± 0.60 |
| "North" | 9.01 ± 1.02 | 1.14 ± 0.43 |
| "South" | 11.89 ± 4.72 | 2.44 ± 1.56 |

Figure 1 presents the energy dependence of the amplitude (a) and phase (b) of the sidereal anisotropy vector projection measured by Andyrchi and BUST together with similar data of other groups [5--13] that made measurements in the energy range from 1 TeV to 1000 TeV. Considerable errors do not allow one to make a definite conclusion about energy dependence of the anisotropy vector projection. Nevertheless, there is some evidence in favor of a small decrease of the magnitude of this projection with increasing energy, and a slight rotation of the vector to smaller phase values.

**Acknowledgments**

The work was partially supported by the State Program of Support for Leading Scientific Schools, grant no. NSh-1828.2003.02.

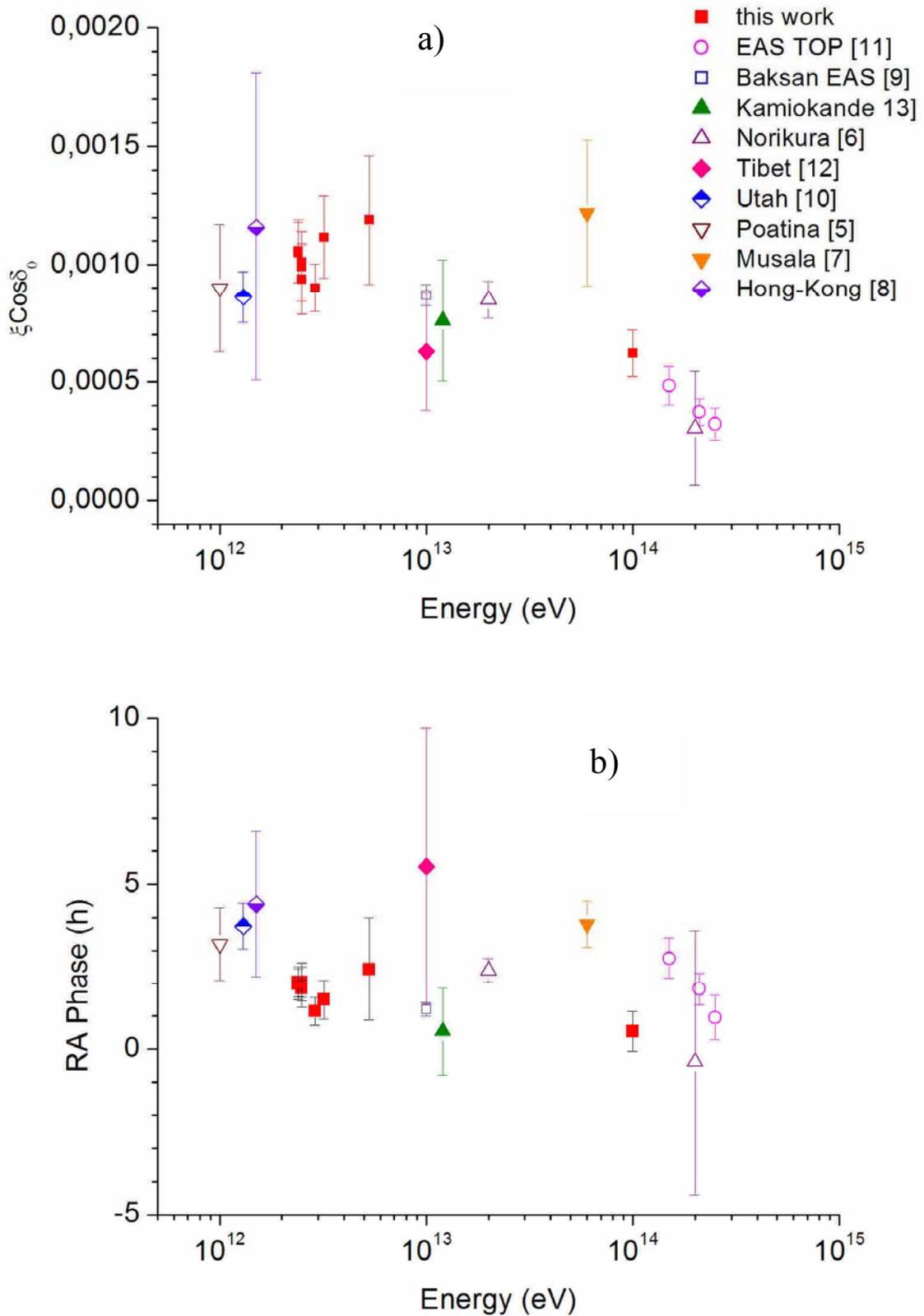

Fig. 1. *Energy dependence of the amplitude (a) and phase (b) of the projections of the sidereal anisotropy vectors onto the equatorial plane according to the data of this work and other experiments ([5-13]).*